\documentclass[11pt,a4paper]{article}
\pdfoutput=1
\usepackage{jheppub}
\usepackage{rotating}
\usepackage{array}
\usepackage{amsmath,amssymb,amstext}
\usepackage[normalem]{ulem}
\usepackage{slashed}
\usepackage{booktabs}
\usepackage[pdftex,table,dvipsnames]{xcolor}
\usepackage{units}
\usepackage{multirow}
\usepackage{url}
\usepackage{color}
\usepackage{xspace}
\usepackage{comment}
\usepackage{enumitem}

\usepackage{multirow}
\renewcommand{\vec}[1]{\mathbf{#1}}

\newcommand{\deta}{\Delta \eta}
\newcommand{\dphi}{\Delta \phi}

\preprint{TTK-23-03, P3H-23-019}

\title{Learning the language of QCD jets with transformers}

\author[a]{Thorben Finke}
\author[a]{\!\!, Michael Kr\"amer}
\author[a]{\!\!, Alexander M\"uck}
\author[b]{and Jan T\"onshoff\;\!}

\affiliation[a]{Institute for Theoretical Particle Physics and Cosmology (TTK),\\ RWTH Aachen University, D-52056 Aachen, Germany}
\affiliation[b]{Chair of Computer Science 7 (Logic and Theory of Discrete Systems), Department of Computer Science, RWTH Aachen University, Aachen, Germany}

\emailAdd{finke@physik.rwth-aachen.de}
\emailAdd{mkraemer@physik.rwth-aachen.de}
\emailAdd{mueck@physik.rwth-aachen.de}
\emailAdd{toenshoff@informatik.rwth-aachen.de}

\abstract{
Transformers have become the primary architecture for natural language processing. In this study, we explore their use for auto-regressive density estimation in high-energy jet physics, which involves working with a high-dimensional space. We draw an analogy between sentences and words in natural language and jets and their constituents in high-energy physics. Specifically, we investigate density estimation for light QCD jets and hadronically decaying boosted top jets. Since transformers allow easy sampling from learned densities, we exploit their generative capability to assess the quality of the density estimate. Our results indicate that the generated data samples closely resemble the original data, as evidenced by the excellent agreement of distributions such as particle multiplicity or jet mass. Furthermore, the generated samples are difficult to distinguish from the original data, even by a powerful supervised classifier. Given their exceptional data processing capabilities, transformers could potentially be trained directly on the massive LHC data sets to learn the probability densities in high-energy jet physics.}

\begin{document}

\maketitle

\section{Introduction}
\label{sec:Introduction}
The Large Hadron Collider (LHC) produces a huge amount of highly structured, high-dimensional data. Estimating the probability density of this data is a crucial task, as it allows one to improve experimental analyses, generate synthetic data, or perform anomaly detection. The LHC data is typically reduced to hand-crafted, high-level features to extract relevant information, a process that is task-dependent and requires expert knowledge. Previous machine learning advances in high-energy physics (HEP) have shown the benefits of using low-level information, which can capture important correlations that may be missed by high-level observables. However, density estimation in many dimensions is inherently challenging due to the curse of dimensionality. 

Density estimation in HEP has primarily focused on generating data samples that follow a given distribution $p(x)$, see the reviews~\cite{Feickert:2021ajf, Butter:2022rso} and references therein. For this purpose, the density need not to be tractable, i.e.\ it is not necessary to be able to calculate $p(x)$ for given $x$. Generative adversarial networks (GANs)~\cite{https://doi.org/10.48550/arxiv.1406.2661} are an example of models that can only sample from a distribution. In contrast, normalizing flows~\cite{https://doi.org/10.48550/arxiv.1505.05770} allow for both sampling and likelihood evaluation. Normalizing flows optimize a bijective mapping with a tractable Jacobian between a simple base distribution and the data distribution, enabling sampling and likelihood evaluation. However, normalizing flows have the disadvantage of a fixed input size due to the bijective mapping, and they are typically only applied to high-level features. While normalizing flows can be applied to high-dimensional physics data, there is a risk of manifold collapse~\cite{https://doi.org/10.48550/arxiv.2204.07172, Cresswell:2022tof}. Nevertheless, refs.~\cite{Krause:2021ilc,Krause:2021wez,Krause:2022jna,Diefenbacher:2023vsw} 
provide examples of successful applications of normalizing flows to such data.

An auto-regressive approach to density estimation was presented in ref.~\cite{Andreassen:2018apy,Andreassen:2019txo}. This approach involves clustering the set of particle momenta within a jet into a binary tree, which is then sequentially analysed by a recurrent neural network. The resulting probability density can be used for discrimination tasks and generating data samples.

We present a novel method for auto-regressive density estimation in jet physics that utilizes the transformer-encoder architecture introduced in ref.\,\cite{vaswani2017attention}. Our approach is closely related to TraDE (Transformers for Density Estimation) introduced in ref.\,\cite{fakoor2020trade}, but we simplify the method to suit our specific needs. Transformers have become the leading architecture in the field of natural language processing (NLP)~\cite{https://doi.org/10.48550/arxiv.1910.03771}. When compared to recurrent neural networks, transformers offer distinct advantages, especially in handling long-range correlations and improving training efficiency. Transformers are now being applied to various HEP tasks, such as jet tagging \cite{Mikuni:2021pou,Qu:2022mxj}, reconstruction \cite{Qiu:2022xvr,DiBello:2022iwf}, learning jet representations \cite{Dillon:2021gag}, and data generation \cite{Kansal:2022spb,Kach:2022uzq,https://doi.org/10.48550/arxiv.2303.05376}. These applications modify the transformer, in particular its attention mechanism, to satisfy desirable physics properties, such as permutational invariance or explicit pairwise interactions. 

We take a different approach and pre-process the data to be more similar to natural language. To represent jet constituents, we discretize their properties by binning their transverse momentum, $p_T$, and the difference in rapidity and azimuthal angle relative to the jet axis, $\deta$ and $\dphi$, respectively. We will argue that this discretization does not result in significant information loss. The discrete particle states can then be viewed as words in a dictionary, and the process of combining particles to form jets is analogous to constructing sentences in natural language processing.

To estimate the density, we use an auto-regressive approach, where the transformer is trained to predict the probability of a jet constituent $p(\vec{x_i})$ given all prior constituents of the jet $\vec{x_1}$ to $\vec{x_{i-1}}$.
In this way, we can model the joint density of the entire jet with $n$ constituents as the product of the probabilities of each constituent
\begin{equation}
    p(\vec{x}) = p(\vec{x_1}) p(\vec{x_2}|\vec{x_1}) \dots p(\vec{x_n}|\vec{x_1}\dots \vec{x_{n-1}}) . \label{eq:prob_chain}
\end{equation}
We introduce a very basic grammar to the language of jets by ordering jet constituents in $p_T$. The transformer setup is flexible, e.g.\ it allows density estimation as well as sampling for jets with varying numbers of constituents. An alternative approach to sampling jets with variable multiplicity using GANs has been presented in ref.~\cite{Buhmann:2023pmh}.

We use Monte Carlo data to train a standard transformer architecture and determine the probability density for QCD and top-jets. By evaluating the likelihood ratio of the densities, we demonstrate that the transformer learns properties specific to the physics of top and QCD jets. To further assess the quality of the density estimation, we generate data samples from the learned probability distributions. The generated samples display excellent agreement with the Monte Carlo data, as demonstrated by distributions such as particle multiplicity or jet mass, and are difficult to distinguish from the data, even for a  powerful supervised classifier. 

This paper is structured as follows: In section~\ref{sec:Setup} we describe the transformer architecture together with the way we pre-process the data. We show results on density estimation as well as sampling in section~\ref{sec:results}. Finally, we conclude the paper in section~\ref{sec:Conclusion} and outline possible future research directions. Additional results are presented in the appendices.

\section{Setup}
\label{sec:Setup}
This chapter provides a detailed description of our setup. Section~\ref{sec:setup_transformer} presents the transformer architecture, the training procedure, and the sampling from the learned probability distribution. Section~\ref{sec:setup_data} introduces the data set and its pre-processing.

\subsection{Transformer encoder for density estimation in high dimensions} \label{sec:setup_transformer}

We use the encoder part of a transformer network as proposed in ref.\,\cite{vaswani2017attention}. As mentioned in the introduction and discussed in detail in section~\ref{sec:setup_data}, the features of each jet constituent are discretized, i.e.\ we consider a large but finite number of different particles, where each $\vec{x}_i$ in eqn.\,(\ref{eq:prob_chain}) is represented by a tuple of integers that describe $p_T$, $\deta$, and $\dphi$. These particles correspond to the words in an NLP dictionary. An embedding layer is used to map the particles back to a trainable continuous representation.

We perform the embedding separately for $p_T$, $\deta$ and $\dphi$. Each embedding consists of a trainable matrix with rows corresponding to the bins in the corresponding feature and a fixed number of columns determined by the embedding dimension. We obtain the final embedding of a particle by summing the three vectors of the embedding dimension. In NLP, this embedding learns to map words with similar meanings close to each other in the embedded space.

The embedding is followed by a block of layers with multi-head self-attention, as implemented in PyTorch~\cite{NEURIPS2019_9015} as \textsc{TransformerEncoderLayer}. The attention mechanism allows the network to learn which positions are especially linked. As in NLP, where words are strongly correlated, the transformer learns the correlations hidden in the jet probability distribution. The multi-head attention allows the network to consider various correlations in a single processing step. This block is followed by a final fully connected layer that maps to the desired output dimension and assigns a probability to each particle of our particle dictionary using a  softmax activation. As we discretized the features, i.e.\ we work with a finite dictionary, we only have to deal with probabilities rather than probability densities. Using Gaussian mixtures, one could  also extend the method to continuous features~\cite{fakoor2020trade}.

We have set the embedding dimension and the output dimension of each \textsc{TransformerEncoderLayer}  to 256. Our model consists of eight such layers, each with four heads. We apply \textsc{LayerNorm} and \textsc{Dropout} of $0.1$ within and after the last \textsc{TransformerEncoderLayer}. With this setup, our network has about 13M trainable parameters. These parameters are optimized for 50 epochs using the Adam optimizer~\cite{Kingma:2014vow} with an initial learning rate of $5\times10^{-4}$, a batch size of 100 and a cosine learning rate schedule with a final learning rate of $10^{-6}$. To maximize the likelihood, we minimize the categorical cross-entropy loss between the predicted next particle and the actual next particle in a jet. Given the exploratory nature of this study, we have not performed hyperparameter optimization. Note that training the transformer model on a NVIDIA Tesla V100 SXM2 GPU takes approximately 5 hours.

During training we use two masks on our data. A fixed length input is desirable to ensure an efficient batch training. Thus, we need to perform padding on jets that have less constituents and mask these padded values using a padding mask. An additional causal mask is used during training to ensure that the network only uses the preceding jet constituents $\vec{x}_1$ to $\vec{x}_{i-1}$ when predicting the current jet constituent $\vec{x}_i$. To technically treat the first jet constituent on the same footing as the others, we add a meaningless zeroth jet constituent $\vec{x}_0$ at the beginning of each jet. For the first particle, the network learns the marginal distribution over the training set. To indicate the end of a jet, we use a stop token as discussed in section~\ref{sec:setup_data}, which allows for the density estimation and the generation of jets with flexible length.

Using the trained network, we obtain the probability $p(\vec{x})$ for a given input jet by multiplying the probabilities $p(\vec{x}_i)$ that the network assigns to the constituents of the jet (including the stop token), according to eqn.\,(\ref{eq:prob_chain}). In practice, we calculate the log-probability and sum up the logarithms of the individual probabilities. For inference, the causal mask allows us to obtain the probability for all particles in the jet at once, with a single forward pass through the network, as was the case during training.

The trained network can also be used to sample jets. However, unlike training and inference, sampling requires an iterative procedure due to the auto-regressive nature of the density estimator, which can impact the speed of sampling. To begin the sampling process, we feed only the zeroth jet constituent into the transformer, and the network provides the probabilities for the first particle. We then sample according to the probabilities, add the sampled particle to the jet, and repeat this procedure until a maximum number of constituents is reached, or the stop token is sampled to indicate that the jet is complete.

As it is well known in NLP, text generation from a transformer network is improved by suppressing the sampling of extremely unlikely possibilities, see for example ref.\,\cite{https://doi.org/10.48550/arxiv.1904.09751}. Here, we adopt this procedure by only sampling from the 5000 most likely particles when predicting the next particle in a jet, which is usually referred to as top-k sampling in NLP. Our choice of $\mathrm{k}=5000$ is motivated at the end of section~\ref{sec:setup_data} after discussing the structure of the data set. Top-k sampling 
reduces the generic problem that the transformer is not able to reduce probabilities for unlikely events to arbitrarily low values. Further discussion on sampling from the full probability distribution without a top-k restriction is provided in appendix~\ref{sec:full_sampling}.

\subsection{Data set and data format} \label{sec:setup_data}

To demonstrate the effectiveness of our method, we use the top tagging reference data set~\cite{kasieczka_gregor_2019_2603256} in this study. This data set comprises 2 milllion  events, split equally between light QCD jets and hadronically decaying top jets. The jets are clustered using the anti-$k_T$ algorithm with $R=0.8$ and selected with $p_T \in [550, 650]$ GeV and $|\eta| < 2$. The data set contains the 200 highest $p_T$ constituents in each jet. We consider only the 50 highest $p_T$ constituents for training the transformer network, but perform density evaluation and sampling with up to 100 constituents to emphasize the flexibility of the setup. We use 600k jets for training and reserve 200k jets each for validation and testing, respectively.

We represent a jet as a list of its constituents sorted by decreasing $p_T$. While alternative orderings inspired by physics, such as those derived from jet clustering algorithms~\cite{Louppe:2017ipp,Andreassen:2018apy,Andreassen:2019txo}, could be employed, we deliberately choose the simple $p_T$ order. This choice is based on the expectation that the transformer model has the ability to learn non-trivial correlations between the jet components without the need for complex task-specific jet setup. For each constituent $i$, we use $p_{T,i}$, $\deta_i = \eta_i - \eta_{\rm jet}$, and $\dphi_i = \phi_i - \phi_{\rm jet}$ as features. We discretize these features by binning. We divide the range $[-0.8, 0.8]$ (set according to the jet radius) into 29 equidistant bins for both $\deta$ and $\dphi$. A similar binning is performed when using jet images. To account for the steeply falling $p_T$ spectrum, we work in log-space and use 39 equidistant bins in the range $[\log(p_T)_{\min}, \log(p_T)_{\max}]$, where $99.9\%$ of all jet constituents in the training data satisfy $p_T>\log(p_T)_{\min}$, and $\log(p_T)_{\max}$ is defined by the constituent with the largest $p_T$. We add a bin 0 as an underflow bin and use bin 40 and 30 as an overflow bin for $p_T$ and $\deta, \dphi$, respectively. Each constituent is then described by three integers corresponding to the respective bins $\hat{p}_T \in [0, 40]$, $\Delta \hat{\eta} \in [0, 30]$, and $\Delta \hat{\phi} \in [0, 30]$. These tuples of integers serve as input for the transformer network.

\begin{figure}
    \centering
    \includegraphics{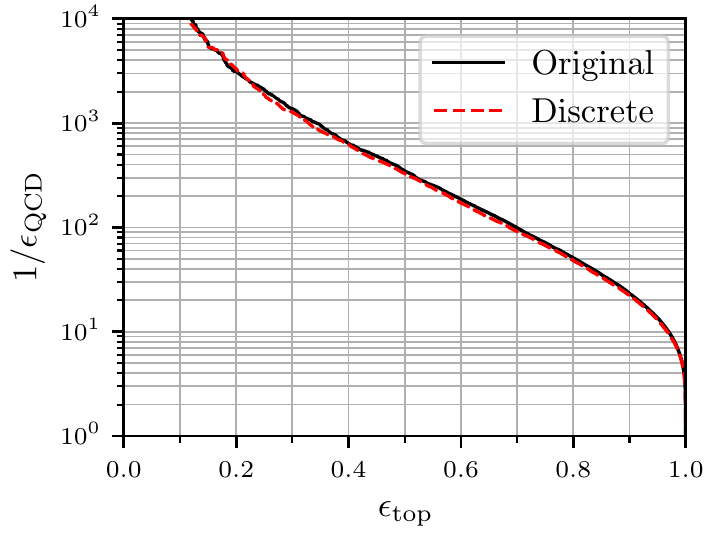}
    \caption{ROC curves in background rejection $1/\epsilon_\mathrm{QCD}$ and signal efficiency $\epsilon_\mathrm{top}$ for top tagging, using the ParticleNet architecture~\cite{Qu:2019gqs} on continuous and discretized data. The data is discretized according to the discription provided in the text. The performance matches the ParticleNet performance stated in ref.\,\cite{Kasieczka:2019dbj}. Discretization does not cause a significant decrease in performance.}
    \label{fig:roc_qcdVStop}
\end{figure}

As described in section~\ref{sec:setup_transformer}, we map these tuples of integers back to a continuous space using an embedding. One could argue that we could instead directly use the continuous values of the features as input. However, we will sample discrete values. These sampled constituents are successively used as input to sample consecutive constituents. Using continuous values during training and discrete values for sampling might reduce the sampling accuracy.

To determine if a significant amount of information is lost using discretized data, we train a ParticleNet classifier~\cite{Qu:2019gqs} to distinguish between top jets and light QCD jets. We excluded features related to the energies of jet constituents or the jet itself, as this information is lost with the features selected for density estimation. The ROC curves for this classifier for both continuous and discrete data, see figure~\ref{fig:roc_qcdVStop}, show that there is indeed no significant loss of information for this task. The same ParticleNet classifier is later used in section~\ref{sec:Sampling} to assess the quality of our generated samples. Specifically, the classifier is tasked with distinguishing between the generated samples and the actual data.

To indicate the end of a jet, we utilize a stop token, which is always a possible prediction for the next particle. To ensure that the network can generate jets with varying numbers of constituents, the stop token is added to the particle dictionary as an additional particle. During training, the stop particle is only added as the last particle of the jet if the jet has less than $N$ constituents, where $N=50$ is the maximum number of training constituents. This approach enables the network to learn what constitutes a complete jet and allows the generation of jets with more than $N$ constituents. The probability of a jet with $n<N$ constituents includes the probability of the stop token appearing at constituent position $n+1$.

\begin{figure}
    \centering
    \includegraphics{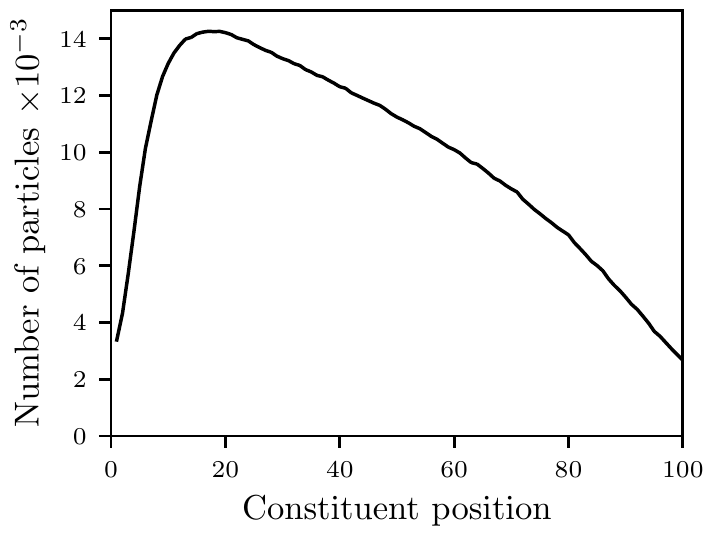}
    \caption{Number of distinct particles that are present at each constituent position.}
    \label{fig:filled_bins}
\end{figure}

Note that not all possible particles are present in the training data; for QCD jets, roughly 35\% of particles are absent from the training set. Figure~\ref{fig:filled_bins} illustrates the number of distinct particles present at each constituent position, with less than half of the possible states being populated even at the peak, and a sharp drop in numbers for positions above 20. Some particles may be present in a more extensive training set, and it is part of the task of density estimation to also extrapolate the density into these regions. However, quite a few particles will actually have zero probability. Sampling from those particles might lead to unsatisfactory results because the transformer cannot suppress the probability of a particle to zero. To avoid this problem we use top-k sampling as introduced at the end of section~\ref{sec:setup_transformer}. To sample mostly from particles that are present in the training set we chose $\mathrm{k}=5000$. We found that varying k by a factor of two did not have a significantly impact on the results presented in section~\ref{sec:results}.

\section{Results}
\label{sec:results}
Employing the setup described in section~\ref{sec:Setup}, the transformer is trained to learn the density of QCD and top jets. Learning the density in a high-dimensional space is challenging, and evaluating the accuracy of the density estimate is equally difficult. In section~\ref{sec:density}, we describe some aspects of our density estimate, specifically the learned jet probabilities. Additionally, in section~\ref{sec:Sampling}, we discuss the generation of jets by sampling from the learned probability distribution to further evaluate the quality of the density estimate.

\subsection{The density estimate} \label{sec:density}

\begin{figure}
    \centering
    \includegraphics{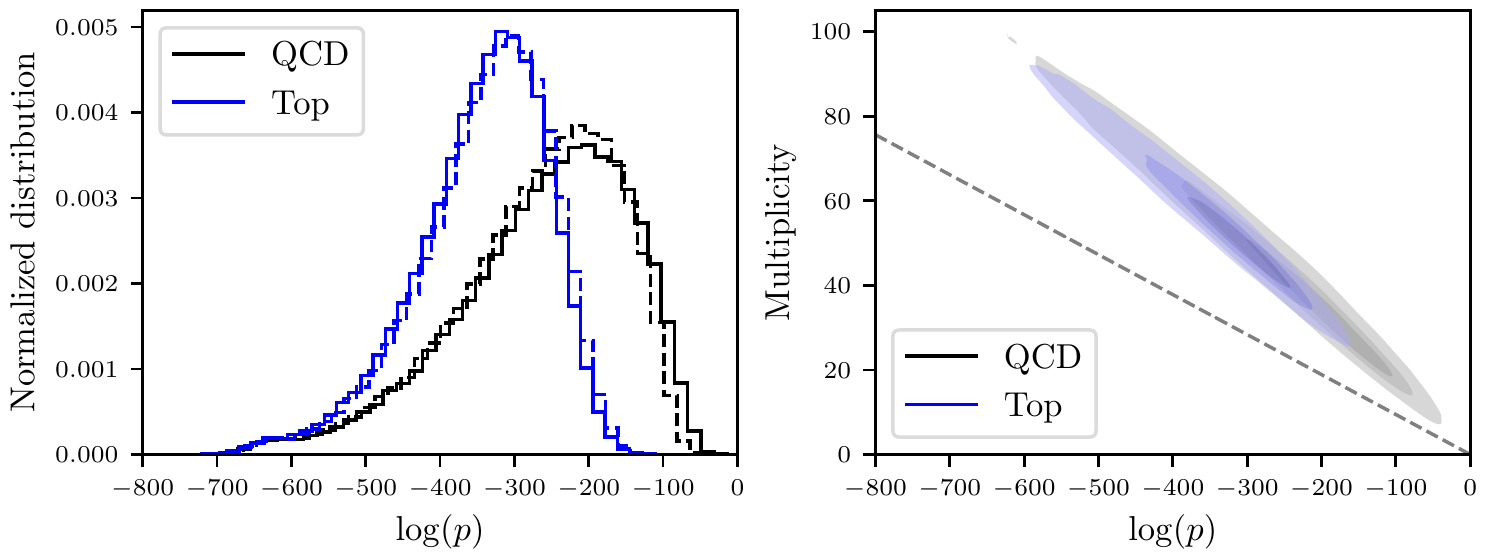}
    \caption{
    Left: Distribution of the estimated jet probability (100 leading constituents) for the QCD jets (black) and the top jets (blue) in the data set. The transformer is either trained on QCD jets (solid) or top jets (dashed). Right: Correlation between the estimated probability and the number of constituents. Contours contain 25\%, 50\% and 75\% of the density using a kernel density estimate (on 50k of the test set). The dashed line shows the jet probability obtained by assigning the same probability to each particle in our dictionary, i.e.\  $\log(p) = \log(1 / N_d) \times n$, where $N_d$ is the size of our particle dictionary and $n$ is the number of constituents for a given jet.
    }
    \label{fig:density}
\end{figure}

The transformer assigns a probability to a given input jet. The distributions of all jet probabilities are shown in figure~\ref{fig:density} (left). While we train on the 50 leading $p_T$ constituents, we show the probability taking into account the 100 leading $p_T$ constituents. As we will see below, the transformer accurately extrapolates to larger jet multiplicities that are not present during training. Note that there is no computational limit to using more constituents for training. We have used 50 constituents to demonstrate the extrapolation capabilities of the sampling. While each individual jet has a small probability, the jet probabilities vary extensively, primarily due to different jet multiplicities.

The correlation between the number of constituents and the probability assigned to a jet is shown in the right panel of figure~\ref{fig:density}. The probability decreases approximately exponentially with increasing number of constituents, because each jet constituent introduces an additional factor in eqn.\,(\ref{eq:prob_chain}). To guide the eye, the probability for a uniform prediction for all particles in our particle dictionary is included in figure~\ref{fig:density}. The decrease in probability for higher multiplicity is (partly) compensated by an increase in the phase space for the jets, i.e.\ in the number of possible jet realizations, and results in the well-known multiplicity distribution to be discussed in section~\ref{sec:Sampling}.

In figure~\ref{fig:density} (left), we also display the probability distribution for top jets, which the network did not encounter during training. The distribution for top jets is shifted toward smaller probabilities, mainly due to their larger multiplicities. As shown in the right panel of figure~\ref{fig:density}, the distribution has approximately the same slope but is shifted to higher multiplicities. The distinction between QCD and top jets based on their underlying physics can be evaluated by training the transformer on top jets (represented by the dashed lines in figure~\ref{fig:density}). As expected, training on top jets, the distribution of probabilities indeed shifts to larger probabilities for the top jets and to smaller probabilities for QCD jets. Although this shift is small compared to the trivial multiplicity offset, it encodes the different physics of the two jet types, since the probabilities for a given jet still change by orders of magnitude.

\begin{figure}
    \centering
    \includegraphics{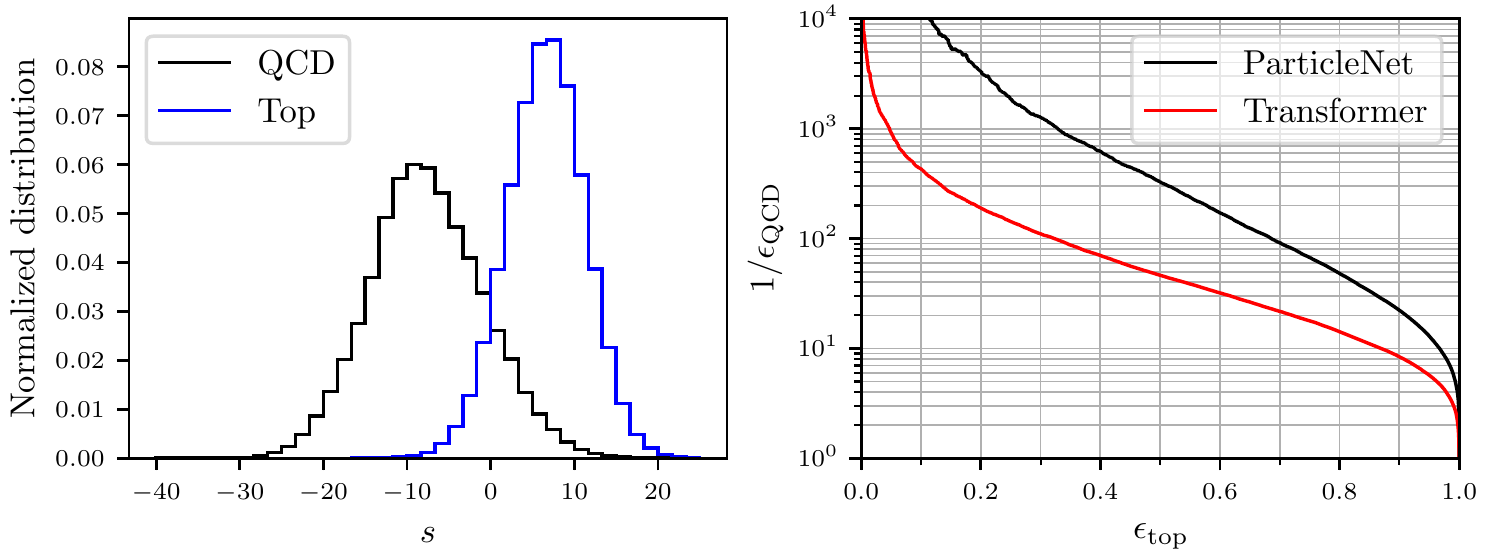}
    \caption{Left: Log-likelihood ratio $s$ for top and QCD jets. Right: ROC curves for top vs.\ QCD classification; the supervised ParticleNet classifier~\cite{Qu:2019gqs} is shown for comparison.}
    \label{fig:density_classification}
\end{figure}

To demonstrate that the transformer not only learns some generic properties but indeed QCD and top specific features, we investigate the classification performance of the likelihood ratio, see also refs.~\cite{Andreassen:2018apy,Andreassen:2019txo,Nachman:2020lpy}. 
We calculate $s = \log(p_t(x)) - \log(p_q(x))$ for any given jet, where $t$ and $q$ represent training on top and QCD jets, respectively. If the densities were perfectly estimated $s$ would be the ideal classification score and monotonically related to the score of an ideal supervised classifier. We show the distribution of $s$ for top and QCD jets in the test set in figure~\ref{fig:density_classification} (left), together with the resulting ROC curve (right). While the likelihood ratio calculated from the low-level features of the jets is not expected to match the performance of a fully supervised classifier like ParticleNet~\cite{Qu:2019gqs}, it still demonstrates some separation capability. Refining the training of the models within the classification framework, as demonstrated in ref.~\cite{Andreassen:2019txo}, could potentially further optimise the discriminative power. In addition, we believe that a significant improvement in the performance of the likelihood-based classifier can be achieved by increasing the amount of training data, as the current limitation appears to be due to overfitting with a training set consisting of only 600k jets.
Notably, evaluating the likelihood-based classifier on the training set even outperforms ParticleNet.

\subsection{Jet generation} \label{sec:Sampling}

In this section, we employ the transformer network as a generative model to sample jets. The transformer network is trained as described in section \ref{sec:Setup} using the 600k light QCD training jets. While the network is only trained on the 50 leading $p_T$ constituents, it samples jets with an arbitrary number of constituents with high fidelity.
Generating jets with up to 100 constituents takes $\sim$20\,ms per jet when sampling from the full distribution. The sampling time is slightly reduced for top-k sampling.\footnote{Timing was tested on a NVIDIA Tesla V100-SXM2-16GB GPU, generating 10k jets with a batch size of 100.}
Where we compare jet samples with our test data, we use discretized values for both.

\begin{figure}
    \centering
    \includegraphics{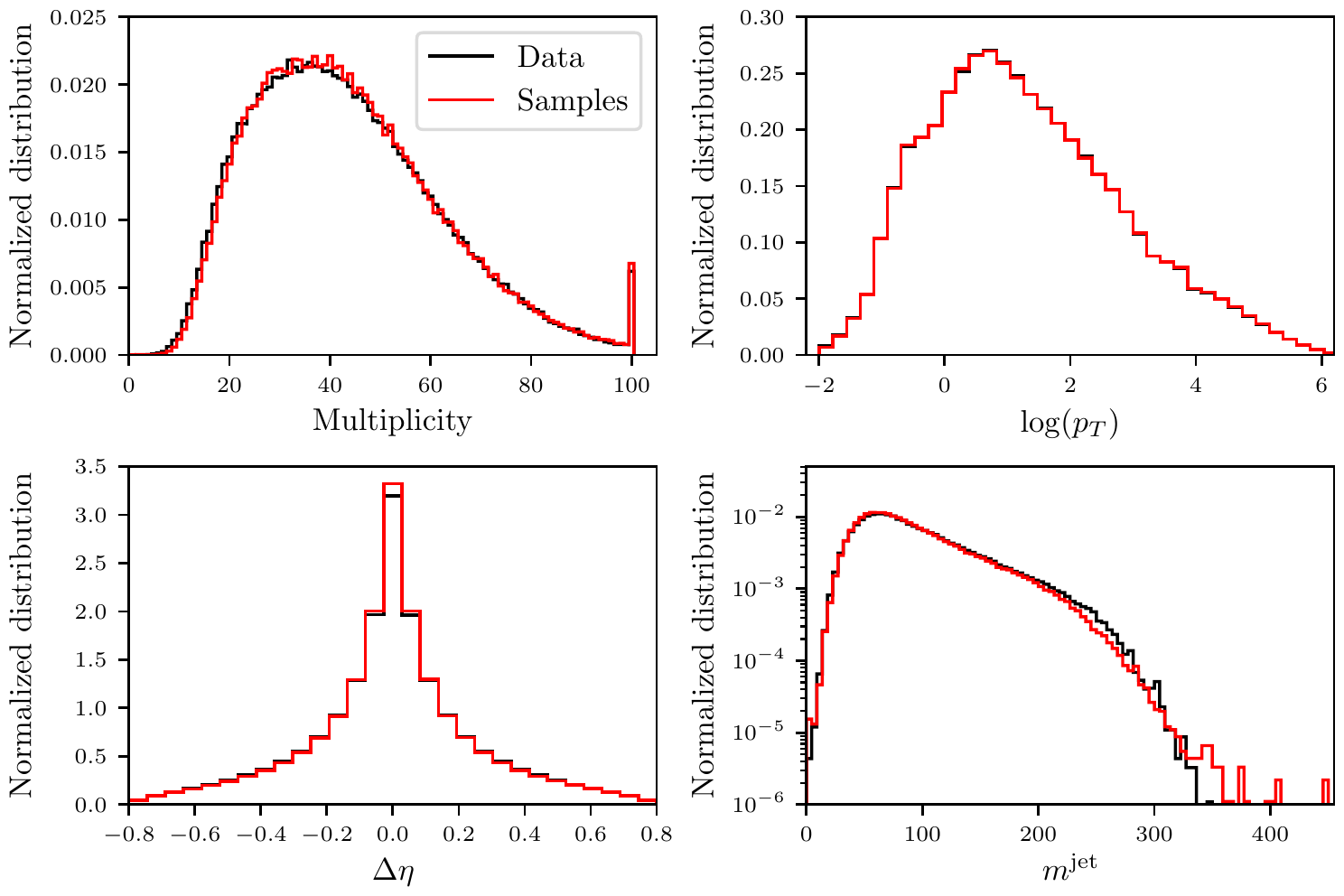}
    \caption{Comparison of data and sample distributions for QCD jets. We compare the particle multiplicity (top left), $\log(p_T)$ (top right), $\deta$ (bottom left), and $m^\mathrm{jet}$ (bottom right) distributions of all jet constituents. We use 200k jets for both the generated samples as well as the data from the test set.
    }
    \label{fig:samples_features}
\end{figure}

In figure~\ref{fig:samples_features}, we compare distributions of 200k sampled events with those from the test set data. The results for the particle multiplicity are particularly interesting. We observe that the transformer does not only learn the multiplicity as seen in the training set where the number of constituents is truncated at 50. The transformer is also able to extrapolate the multiplicity to larger values. Hence, it learns the structure of jets to the extent that it knows how to terminate them. 

In figure~\ref{fig:samples_features}, we also present the distributions of the input features $p_T$ and $\deta$, along with the jet mass calculated assuming massless constituents and the central values for the input features in each bin. We do not show the distribution of $\dphi$ for brevity since it closely resembles that of $\deta$. The $p_T$ distribution of all constituents is accurately reproduced, and the jet mass is well-reproduced. Although the angular distributions exhibit some minor deviations, with the sampling being slightly too central on average for QCD jets, the generated samples show an impressive overall agreement with the data.

\begin{figure}
    \centering
    \includegraphics{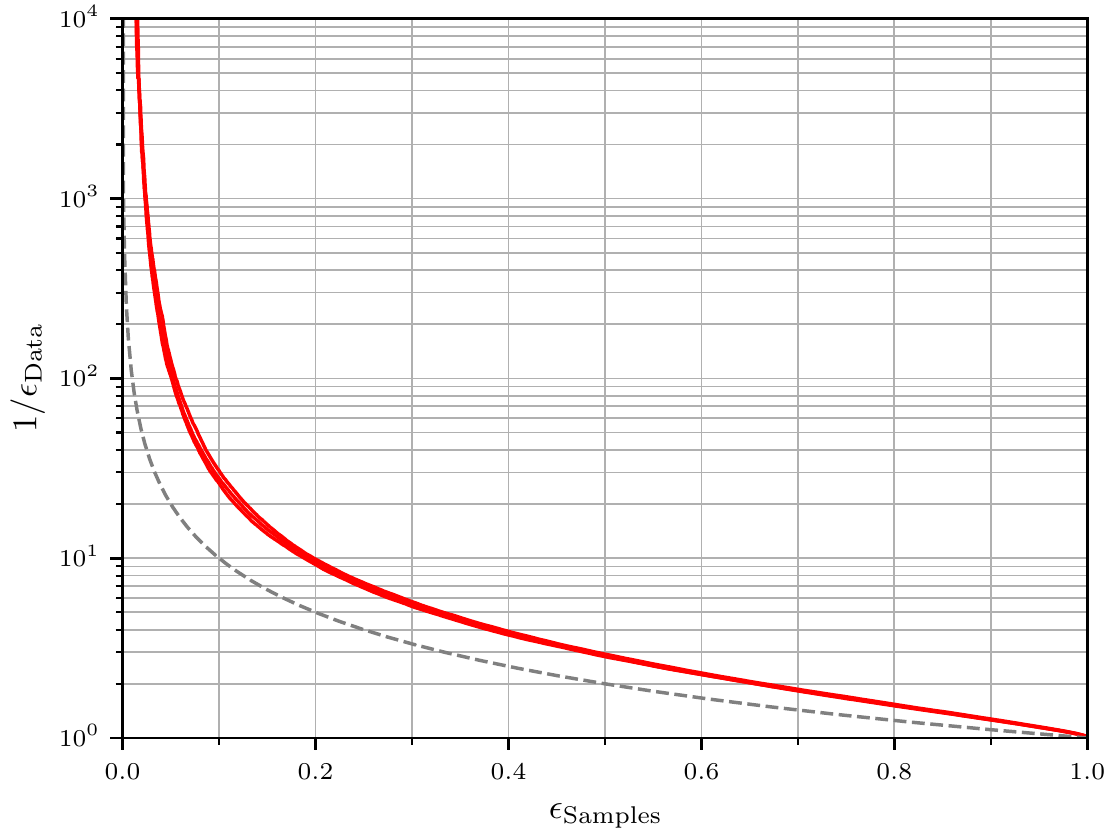}
    \caption{ROC curves for a ParticleNet classifier to distinguish 200k generated samples from the 200k  jets of the test data set. We display three separate curves, each corresponding to independent training, sampling, and classification. The dashed line shows the ROC curve of a random classifier.}
    \label{fig:samples_roc}
\end{figure}

To assess the quality of our generated samples beyond one-dimensional distributions, we train a ParticleNet classifier (see section~\ref{sec:Setup}) to differentiate between the samples and the data. This is an important test because generative models often produce samples that are easily distinguishable from the real data by a classifier (see for example the discussion in refs.~\cite{Krause:2021ilc, Diefenbacher:2023vsw}). The resulting ROC curves, shown in figure~\ref{fig:samples_roc}, indicate that our generative model produces samples of high fidelity. We display three separate curves, each corresponding to independent training, sampling, and classification. Despite the ParticleNet classifier's high discriminative power, its AUC score of 0.62 is only slightly better than random. The success of our generative model in passing this test can be attributed to its excellent density estimation capabilities. Results for sampling without applying the top-k method are discussed in appendix~\ref{sec:full_sampling}. Results for evaluating the quality of the density estimation for top jets by sampling can be found in appendix~\ref{sec:results_top}.

\section{Conclusion}
\label{sec:Conclusion}
Transformers are extremely powerful deep learning architectures that have set new standards in natural language processing. In particular, large and powerful models can be trained if sufficient amounts of data are available. In this study, we investigate the use of transformers for determining probability densities in jet physics. We consider low-level information, such as the transverse momentum and angular position of the jet constituents, rather than relying on hand-crafted high-level features. While low-level information can capture important correlations in the data, density estimation in such a high-dimensional space remains a challenging task. 

We draw an analogy between sentences and words in natural language and jets and their constituents in high-energy physics. To make the data more similar to natural language, we pre-process the features of the jet constituents by discretizing them. We have shown that this discretization does not lead to a significant loss of information. The discrete particle states can then be viewed as words in a dictionary, and the process of combining particles into jets is analogous to constructing sentences in natural language processing.

We use an auto-regressive approach to determine the probability density of QCD and top-jets. In this approach, the transformer is trained to determine the probability of a jet constituent given all the previous constituents of the jet. The design of the transformer is flexible, allowing for both density estimation and the generation of artificial jet samples with different numbers of constituents.

We train a standard transformer architecture on 600k Monte Carlo data to determine the probability density for QCD and top-jets. By evaluating the likelihood ratio of the densities for QCD and top-jets, we demonstrate that the transformer learns properties specific to the physics of top and QCD jets. To assess the quality of the density estimation in more detail, we use the transformer to generate data samples from the learned probability distributions. To suppress the sampling of extremely unlikely particle configurations, we utilize top-k sampling, a method commonly used in natural language processing.

The generated data samples exhibit impressive overall agreement with the data, as demonstrated by the excellent agreement of distributions such as particle multiplicity or jet mass. Additionally, the generated samples are difficult to distinguish from the data, even for a highly powerful supervised classifier, a feature which is hard to achieve with other generative models on low-level LHC data.

Our analysis has shown that transformers can effectively learn the probability densities of jets. Given their ability to process vast amounts of data, larger and more powerful transformers can be trained on the enormous data sets generated so far at the LHC and expected in the future.

Moving forward, also our specific study on jets can be extended in several directions. One interesting avenue is investigating how the transformer architecture can be generalized to continuous variables,  either by turning to Gaussian mixture models or even in a more straightforward way by increasing the particle dictionary to a size where the binning is practically irrelevant. Another promising direction is optimizing the hyperparameters and increasing the training data to further improve the performance of the model. We plan to pursue these questions in a future publication.

\section*{Acknowledgements}
We would like to thank Martin Grohe for discussions and Erik Buhmann, Gregor Kasieczka and David Shih for valuable comments and suggestions on the draft paper. TF is supported by the Deutsche Forschungsgemeinschaft (DFG, German Research Foundation) under grant 400140256 - GRK 2497: The physics of the heaviest particles at the Large Hadron Collider. The research of MK and AM is supported by the Deutsche Forschungsgemeinschaft (DFG, German Research Foundation) under grant 396021762 - TRR 257 “Particle Physics Phenomenology after the Higgs Discovery”. JT is supported by the Deutsche Forschungsgemeinschaft (DFG, German Research Foundation) under grants GR 1492/16-1 and KI 2348/1-1 “Quantitative Reasoning About Database Queries”. The authors gratefully acknowledge the computing time granted by the NHR4CES Resource Allocation Board and provided on the supercomputer CLAIX at RWTH Aachen University as part of the NHR4CES infrastructure. The calculations for this research were conducted with computing resources under the project rwth0934. 

\begin{appendix}
\section{Further results}
\label{sec:more_results}
\subsection{Sampling QCD jets from the full distribution}
\label{sec:full_sampling}

\begin{figure}
    \centering
    \includegraphics{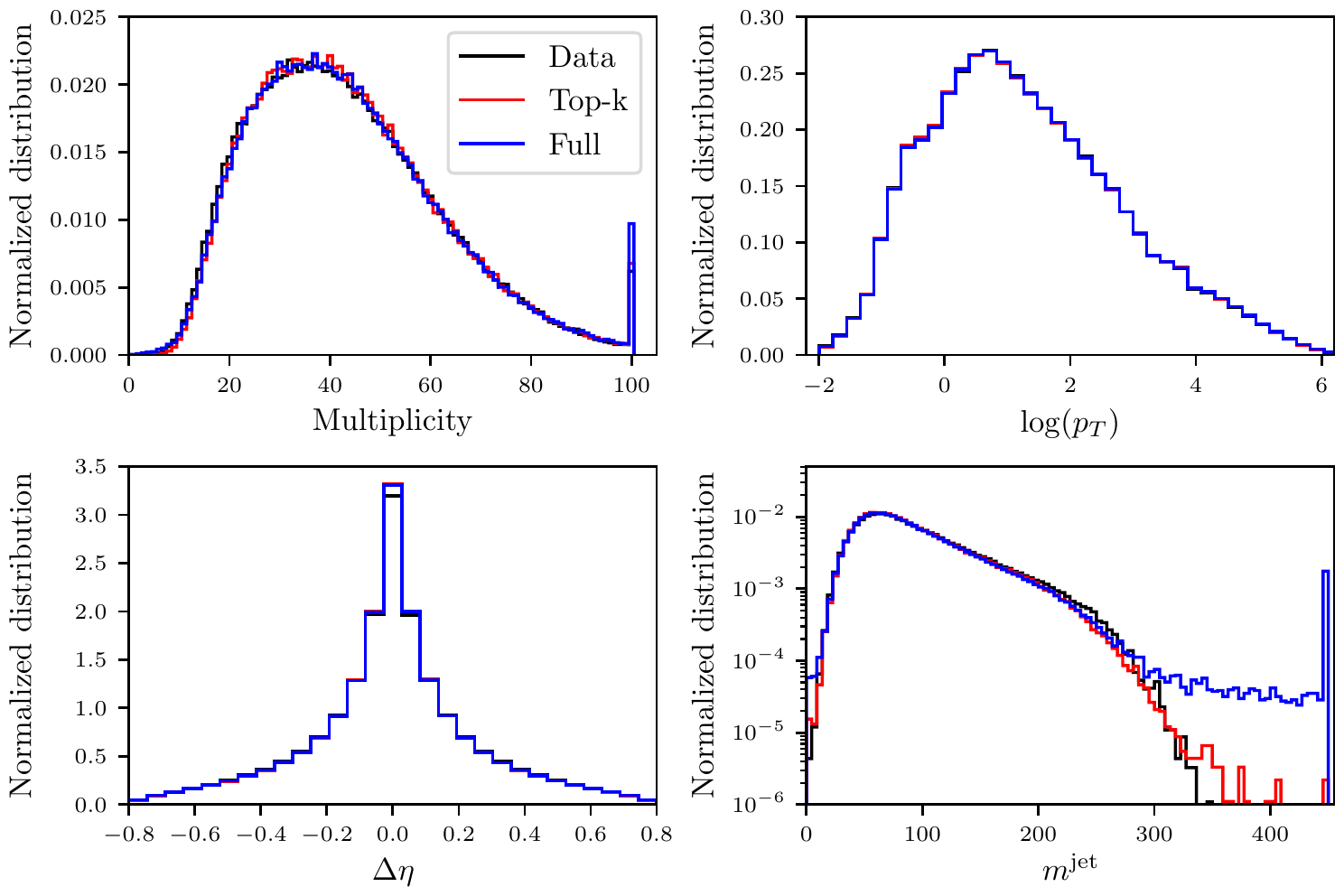}
    \caption{Comparison of data and sample distributions for QCD jets. As in figure~\ref{fig:samples_features}, but now including the distributions obtained from sampling the full probability distribution (blue curve).}
    \label{fig:qcd_full_samples}
\end{figure}

In the main part of the paper, we present the results of sampling using the top-k method, with a value of $\mathrm{k}=5000$. For more information, see section~\ref{sec:setup_transformer}. In figure~\ref{fig:qcd_full_samples}, we show the results when sampling from the full probability distribution, as estimated by the transformer. While the multiplicity and transverse momentum distributions, as well as the angular distributions, are almost identical to those obtained through top-k sampling, there is a slight oversampling of the multiplicity distribution in the overflow bin. Moreover, we observe an unphysical tail at high jet masses, indicating that the transformer is oversampling very unlikely particle configurations that it cannot adequately suppress.

\begin{figure}
    \centering
    \includegraphics{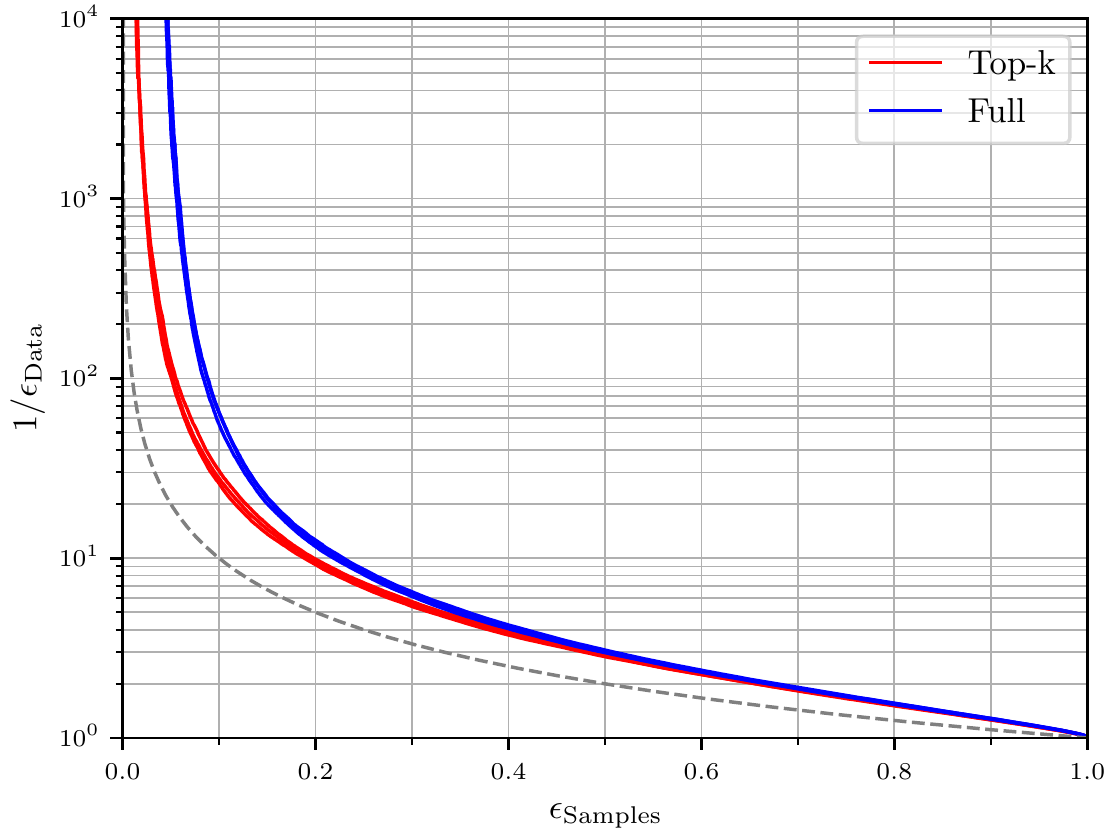}
    \caption{ROC curves for a ParticleNet classifier to distinguish 200k generated QCD jet samples from the
200k jets of the test data set. As in figure~\ref{fig:samples_roc}, but now including the ROC curves obtained from sampling the full probability distribution (blue curve).}
    \label{fig:qcd_full_roc}
\end{figure}

In figure~\ref{fig:qcd_full_roc}, we show the ROC curve obtained through a ParticleNet classifier used to differentiate between the samples from the full probability distribution and the data. While the ROC curve appears similar to that of the top-5k samples for large $\epsilon_\mathrm{sample}$, there is a slightly larger fraction (about 5\%) of sampled events that can be easily separated from the data. With top-k sampling this fraction can be reduced to the 1\% level, because less unphysical jets are sampled from the low-probability region where the transformer struggles to suppress these probabilities efficiently. 

\subsection{Sampling top jets}
\label{sec:results_top}

\begin{figure}
    \centering
    \includegraphics{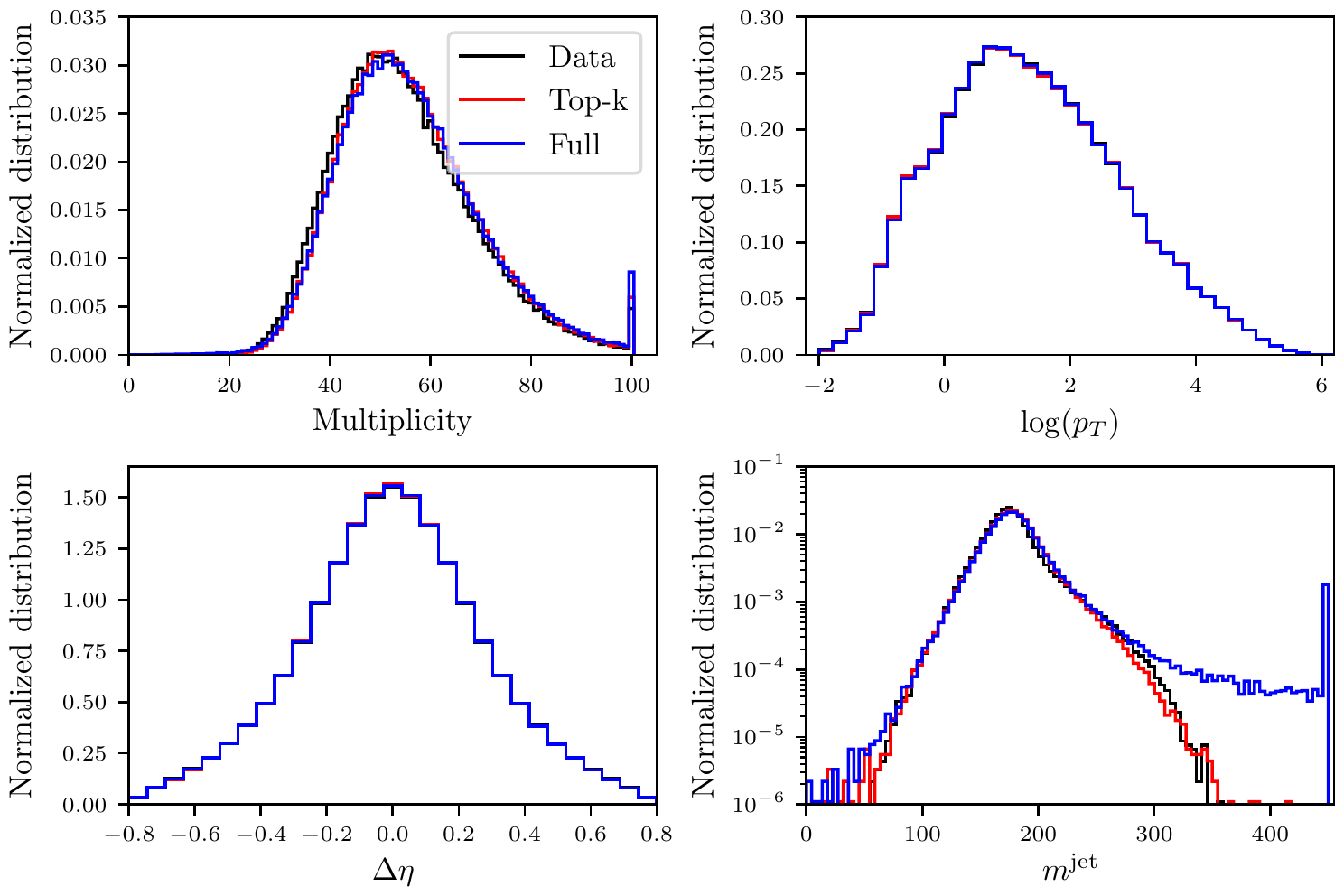}
    \caption{omparison of data and sample distributions in jet multiplicity, $\log(p_T)$, $\deta$ and the jet mass $m^\mathrm{jet}$ for top jets. We show the distributions obtained from top-k and full sampling.}
    \label{fig:top_full_samples}
\end{figure}

In figure~\ref{fig:top_full_samples}, we present distributions of various observables for sampled top jets. Compared to QCD jets, the multiplicity distribution of the data is somewhat less well-reproduced for top jets. This is perhaps not surprising, as our training only includes 50 constituents, which captures only a fraction of the complexity of a typical top jet. However, the one-dimensional distributions in $\log(p_T)$, $\deta$ and $m^\mathrm{jet}$ are well-reproduced, with a level of agreement between the samples and data similar to that for QCD jets.

\begin{figure}
    \centering
    \includegraphics{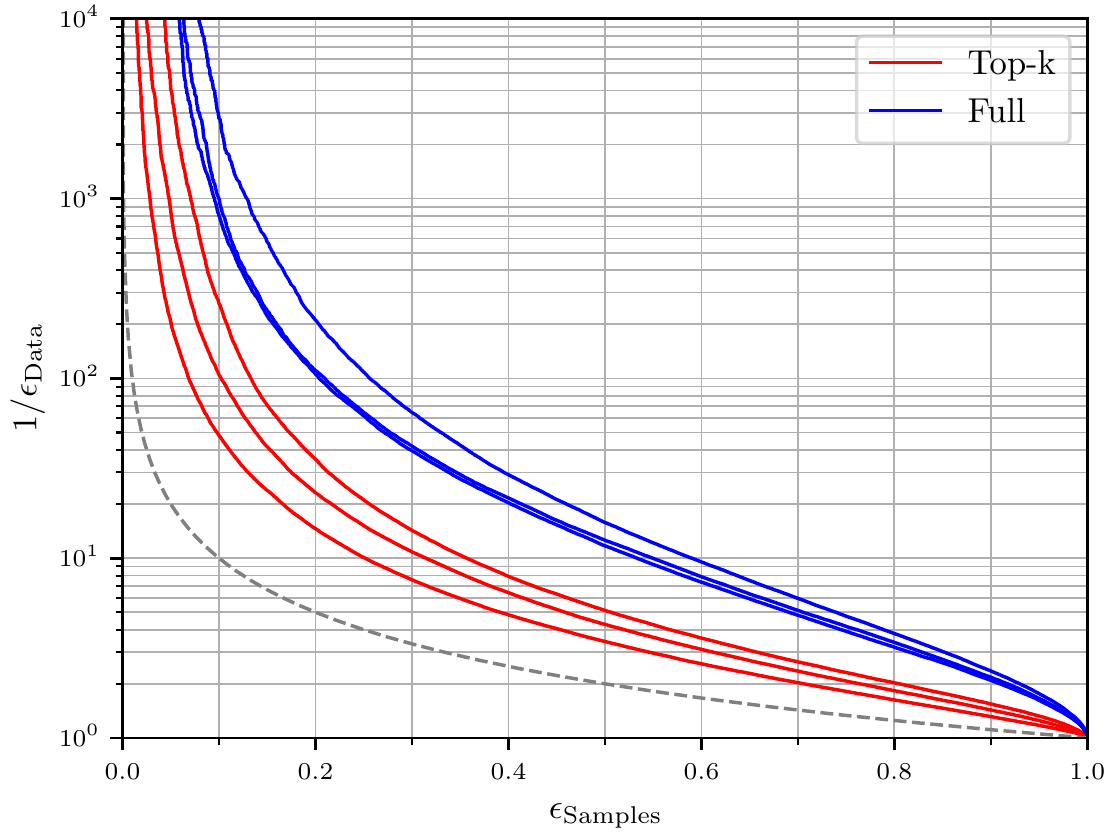}
    \caption{ROC curves for a ParticleNet classifier to distinguish 200k generated top jet samples from the 200k jets of the test data set. We show three ROC curves corresponding to independent training, sampling, and classification for top-k and full sampling, respectively.}
    \label{fig:top_full_roc}
\end{figure}

Figure~\ref{fig:top_full_roc} shows the ROC curves for classification between the top jet samples and the data. Compared to figure~\ref{fig:qcd_full_roc} for QCD jets, we observe a slightly better separation between the samples and the data for top jets, with an AUC score of about 0.7 for the top-k results. We also observe a wider spread in the results obtained by independent training, sampling and classification, and a stronger influence of sampling from the 5k most probable particles compared to QCD jets. These results suggest that more extensive training on a larger data set is required to further improve the probability density of top jets, which have an inherently more complex structure than QCD jets. We will address the need for more comprehensive training, including a larger number of constituents and a much larger training data set in future research.

\end{appendix}

\newpage
\bibliographystyle{JHEP_improved}
\bibliography{bibliography.bib}

\providecommand{\href}[2]{#2}\begingroup\raggedright\begin{thebibliography}{10}

\bibitem{Feickert:2021ajf}
M.~Feickert and B.~Nachman, {\it {A Living Review of Machine Learning for
  Particle Physics}},  \href{http://arxiv.org/abs/2102.02770}{{\tt
  2102.02770}}.

\bibitem{Butter:2022rso}
S.~Badger et~al., {\it {Machine Learning and LHC Event Generation}},
  \href{http://arxiv.org/abs/2203.07460}{{\tt 2203.07460}}.

\bibitem{https://doi.org/10.48550/arxiv.1406.2661}
I.~J. Goodfellow, J.~Pouget-Abadie, M.~Mirza, B.~Xu, D.~Warde-Farley, et~al.,
  {\it {Generative Adversarial Networks}},
  \href{http://arxiv.org/abs/1406.2661}{{\tt 1406.2661}}.

\bibitem{https://doi.org/10.48550/arxiv.1505.05770}
D.~J. Rezende and S.~Mohamed, {\it {Variational Inference with Normalizing
  Flows}},  \href{http://arxiv.org/abs/1505.05770}{{\tt 1505.05770}}.

\bibitem{https://doi.org/10.48550/arxiv.2204.07172}
G.~Loaiza-Ganem, B.~L. Ross, J.~C. Cresswell, and A.~L. Caterini, {\it
  {Diagnosing and Fixing Manifold Overfitting in Deep Generative Models}},
  \href{http://arxiv.org/abs/2204.07172}{{\tt 2204.07172}}.

\bibitem{Cresswell:2022tof}
J.~C. Cresswell, B.~L. Ross, G.~Loaiza-Ganem, H.~Reyes-Gonzalez, M.~Letizia,
  et~al., {\it {CaloMan: Fast generation of calorimeter showers with density
  estimation on learned manifolds}},  in {\em {36th Conference on Neural
  Information Processing Systems}}, 11, 2022.
\newblock \href{http://arxiv.org/abs/2211.15380}{{\tt 2211.15380}}.

\bibitem{Krause:2021ilc}
C.~Krause and D.~Shih, {\it {CaloFlow: Fast and Accurate Generation of
  Calorimeter Showers with Normalizing Flows}},
  \href{http://arxiv.org/abs/2106.05285}{{\tt 2106.05285}}.

\bibitem{Krause:2021wez}
C.~Krause and D.~Shih, {\it {CaloFlow II: Even Faster and Still Accurate
  Generation of Calorimeter Showers with Normalizing Flows}},
  \href{http://arxiv.org/abs/2110.11377}{{\tt 2110.11377}}.

\bibitem{Krause:2022jna}
C.~Krause, I.~Pang, and D.~Shih, {\it {CaloFlow for CaloChallenge Dataset 1}},
  \href{http://arxiv.org/abs/2210.14245}{{\tt 2210.14245}}.

\bibitem{Diefenbacher:2023vsw}
S.~Diefenbacher, E.~Eren, F.~Gaede, G.~Kasieczka, C.~Krause, et~al., {\it
  {L2LFlows: Generating High-Fidelity 3D Calorimeter Images}},
  \href{http://arxiv.org/abs/2302.11594}{{\tt 2302.11594}}.

\bibitem{Andreassen:2018apy}
A.~Andreassen, I.~Feige, C.~Frye, and M.~D. Schwartz,
  \href{http://dx.doi.org/10.1140/epjc/s10052-019-6607-9}{{\it {JUNIPR: a
  Framework for Unsupervised Machine Learning in Particle Physics}}, } {\em
  Eur. Phys. J. C} {\bf 79} (2019), no.~2 102,
  [\href{http://arxiv.org/abs/1804.09720}{{\tt 1804.09720}}].

\bibitem{Andreassen:2019txo}
A.~Andreassen, I.~Feige, C.~Frye, and M.~D. Schwartz,
  \href{http://dx.doi.org/10.1103/PhysRevLett.123.182001}{{\it {Binary JUNIPR:
  an interpretable probabilistic model for discrimination}}, } {\em Phys. Rev.
  Lett.} {\bf 123} (2019), no.~18 182001,
  [\href{http://arxiv.org/abs/1906.10137}{{\tt 1906.10137}}].

\bibitem{vaswani2017attention}
A.~Vaswani, N.~Shazeer, N.~Parmar, J.~Uszkoreit, L.~Jones, et~al., {\it
  {Attention is all you need}},  {\em Advances in neural information processing
  systems} {\bf 30} (2017) [\href{http://arxiv.org/abs/1706.03762}{{\tt
  1706.03762}}].

\bibitem{fakoor2020trade}
R.~Fakoor, P.~Chaudhari, J.~Mueller, and A.~J. Smola, {\it {TraDE: Transformers
  for Density Estimation}},  \href{http://arxiv.org/abs/2004.02441}{{\tt
  2004.02441}}.

\bibitem{https://doi.org/10.48550/arxiv.1910.03771}
T.~Wolf, L.~Debut, V.~Sanh, J.~Chaumond, C.~Delangue, et~al., {\it
  {HuggingFace's Transformers: State-of-the-art Natural Language Processing}},
  \href{http://arxiv.org/abs/1910.03771}{{\tt 1910.03771}}.

\bibitem{Mikuni:2021pou}
V.~Mikuni and F.~Canelli,
  \href{http://dx.doi.org/10.1088/2632-2153/ac07f6}{{\it {Point cloud
  transformers applied to collider physics}}, } {\em Mach. Learn. Sci. Tech.}
  {\bf 2} (2021), no.~3 035027, [\href{http://arxiv.org/abs/2102.05073}{{\tt
  2102.05073}}].

\bibitem{Qu:2022mxj}
H.~Qu, C.~Li, and S.~Qian, {\it {Particle Transformer for Jet Tagging}},
  \href{http://arxiv.org/abs/2202.03772}{{\tt 2202.03772}}.

\bibitem{Qiu:2022xvr}
S.~Qiu, S.~Han, X.~Ju, B.~Nachman, and H.~Wang, {\it {A Holistic Approach to
  Predicting Top Quark Kinematic Properties with the Covariant Particle
  Transformer}},  \href{http://arxiv.org/abs/2203.05687}{{\tt 2203.05687}}.

\bibitem{DiBello:2022iwf}
F.~A. Di~Bello et~al., {\it {Reconstructing particles in jets using set
  transformer and hypergraph prediction networks}},
  \href{http://arxiv.org/abs/2212.01328}{{\tt 2212.01328}}.

\bibitem{Dillon:2021gag}
B.~M. Dillon, G.~Kasieczka, H.~Olischlager, T.~Plehn, P.~Sorrenson, et~al.,
  \href{http://dx.doi.org/10.21468/SciPostPhys.12.6.188}{{\it {Symmetries,
  safety, and self-supervision}}, } {\em SciPost Phys.} {\bf 12} (2022), no.~6
  188, [\href{http://arxiv.org/abs/2108.04253}{{\tt 2108.04253}}].

\bibitem{Kansal:2022spb}
R.~Kansal, A.~Li, J.~Duarte, N.~Chernyavskaya, M.~Pierini, et~al., {\it {On the
  Evaluation of Generative Models in High Energy Physics}},
  \href{http://arxiv.org/abs/2211.10295}{{\tt 2211.10295}}.

\bibitem{Kach:2022uzq}
B.~K\"ach, D.~Kr\"ucker, and I.~Melzer-Pellmann, {\it {Point Cloud Generation
  using Transformer Encoders and Normalising Flows}},
  \href{http://arxiv.org/abs/2211.13623}{{\tt 2211.13623}}.

\bibitem{https://doi.org/10.48550/arxiv.2303.05376}
M.~Leigh, D.~Sengupta, G.~Quétant, J.~A. Raine, K.~Zoch, et~al., {\it
  {PC-JeDi: Diffusion for Particle Cloud Generation in High Energy Physics}},
  \href{http://arxiv.org/abs/2303.05376}{{\tt 2303.05376}}.

\bibitem{Buhmann:2023pmh}
E.~Buhmann, G.~Kasieczka, and J.~Thaler, {\it {EPiC-GAN: Equivariant Point
  Cloud Generation for Particle Jets}},
  \href{http://arxiv.org/abs/2301.08128}{{\tt 2301.08128}}.

\bibitem{NEURIPS2019_9015}
A.~Paszke, S.~Gross, F.~Massa, A.~Lerer, J.~Bradbury, et~al.,
  \href{http://papers.neurips.cc/paper/9015-pytorch-an-imperative-style-high-performance-deep-learning-library.pdf}{{\it
  {PyTorch: An Imperative Style, High-Performance Deep Learning Library}}, } in
  {\em Advances in Neural Information Processing Systems 32}, pp.~8024--8035.
\newblock Curran Associates, Inc., 2019.

\bibitem{Kingma:2014vow}
D.~P. Kingma and J.~Ba, {\it {Adam: A Method for Stochastic Optimization}},
  \href{http://arxiv.org/abs/1412.6980}{{\tt 1412.6980}}.

\bibitem{https://doi.org/10.48550/arxiv.1904.09751}
A.~Holtzman, J.~Buys, L.~Du, M.~Forbes, and Y.~Choi, {\it {The Curious Case of
  Neural Text Degeneration}},  \href{http://arxiv.org/abs/1904.09751}{{\tt
  1904.09751}}.

\bibitem{kasieczka_gregor_2019_2603256}
G.~Kasieczka, T.~Plehn, J.~Thompson, and M.~Russel,
  \href{https://doi.org/10.5281/zenodo.2603256}{{\it {Top Quark Tagging
  Reference Dataset}}, } Mar., 2019.

\bibitem{Louppe:2017ipp}
G.~Louppe, K.~Cho, C.~Becot, and K.~Cranmer,
  \href{http://dx.doi.org/10.1007/JHEP01(2019)057}{{\it {QCD-Aware Recursive
  Neural Networks for Jet Physics}}, } {\em JHEP} {\bf 01} (2019) 057,
  [\href{http://arxiv.org/abs/1702.00748}{{\tt 1702.00748}}].

\bibitem{Qu:2019gqs}
H.~Qu and L.~Gouskos, \href{http://dx.doi.org/10.1103/PhysRevD.101.056019}{{\it
  {ParticleNet: Jet Tagging via Particle Clouds}}, } {\em Phys. Rev. D} {\bf
  101} (2020), no.~5 056019, [\href{http://arxiv.org/abs/1902.08570}{{\tt
  1902.08570}}].

\bibitem{Kasieczka:2019dbj}
A.~Butter et~al., \href{http://dx.doi.org/10.21468/SciPostPhys.7.1.014}{{\it
  {The Machine Learning landscape of top taggers}}, } {\em SciPost Phys.} {\bf
  7} (2019) 014, [\href{http://arxiv.org/abs/1902.09914}{{\tt 1902.09914}}].

\bibitem{Nachman:2020lpy}
B.~Nachman and D.~Shih,
  \href{http://dx.doi.org/10.1103/PhysRevD.101.075042}{{\it {Anomaly Detection
  with Density Estimation}}, } {\em Phys. Rev. D} {\bf 101} (2020) 075042,
  [\href{http://arxiv.org/abs/2001.04990}{{\tt 2001.04990}}].

\end{thebibliography}\endgroup

\end{document}